\documentclass{article}
\usepackage{iclr2026_conference,times}
\usepackage{graphicx}
\usepackage{subcaption}
\usepackage{epstopdf}
\usepackage{amsmath,amsfonts,amssymb}
\usepackage{hyperref}
\usepackage{url}
\usepackage{xcolor}
\usepackage{skak}
\usepackage{chessboard}
\hypersetup{colorlinks=true,citecolor=blue,urlcolor=blue,linkcolor=blue}

\title{Three-Body Alignment: Aligning Chess Agent with Human Reasoning through Reranked Rationale}

\author{
  Jaymari Chua$^{1}$, Chen Wang$^{2}$, Liming Zhu$^{2}$, Lina Yao$^{1}$ \\
  \vspace{0.15cm} \\
  $^{1}$University of New South Wales, UNSW Sydney, Kensington, NSW 2052, Australia \\
  \texttt{\{jace.chua,lina.yao\}@unsw.edu.au} \\
  \texttt{ORCID: 0009-0008-4900-2660, 0000-0002-4149-839X} \\
  \vspace{0.15cm} \\
  $^{2}$CSIRO's Data61, Australian Technology Park, 13 Garden Street, Eveleigh, NSW 2015 \\
  \texttt{\{chen.wang,liming.zhu\}@data61.csiro.au} \\
  \texttt{ORCID: 0000-0002-3119-4763, 0000-0001-5839-3765}
}

\iclrfinalcopy
\begin{document}

\maketitle

\begin{abstract}
As reasoning agents become increasingly complex, aligning their underlying reasoning and decision-making processes with human conceptual models is a challenge for AI security and safety. When modelling expert knowledge, understanding how to characterise and integrate insights from agents with fundamentally different reasoning architectures is necessary for safe and predictable deployment. We investigate this alignment through a \emph{three-body alignment} in chess, analysing the semantic divergence between rationales produced by human experts (Grandmasters), engine-assisted human commentators (who rationalise the outputs of efficiently updatable neural networks, or NNUEs), and Large Language Models (LLMs). Our contributions include: (1) A novel multisource rationale dataset, constructed using an agentic data engineering pipeline to transform unstructured expert commentary into structured, queryable data for alignment evaluation. (2) An empirical analysis of the semantic embedding space. Using t-SNE visualisation, we demonstrate that these sources form distinct clusters, confirming significant heterogeneity and reflecting fundamentally different conceptual approaches to the same environment. (3) An experiment demonstrating that reranking mechanisms can improve human alignment, while quantifying the explicit trade-off with tactical performance, offering a pathway for more interpretable agent decision-making. (4) The preliminary development of an enriched chess narrative dataset structure, designed to lay the groundwork for future evaluations of text rationale similarity and to address the limitations of standard dense retrieval. (5) Finally, we open-source our chess rationales dataset\footnote{Hugging Face: \url{https://huggingface.co/datasets/jaymarichua/trichess}} to support developing novel techniques that integrate diverse expert knowledge into human-aligned intelligent agents.
\end{abstract}

\paragraph{Keywords:} Large Language Models, AI Alignment, Explainable AI (XAI), Information Retrieval, Security Alignment

\section{Introduction}

For decades, the game of chess has served as a testbed for measuring progress in artificial intelligence. Yet the progress of AI in this space has largely prioritised computational prowess over human understanding. Traditional machine learning systems, heavily optimised for next-move prediction \cite{tang2024maia,ruoss2024amortized,zhang2025complete}, frequently generate solutions derived from multi-depth calculations. While predominating techniques that rely on exhaustive computation are strategically sound for the domain, the generated solutions are often difficult for human players to interpret and differ significantly from how human players tend to make decisions \cite{tang2024maia,zhang2024human}.

Initiatives such as MAIA specifically target this human-AI mismatch by training models to emulate human behaviour across various skill levels \cite{tang2024maia}. In addition, Large Language Models (LLMs) and specialised Large Reasoning Models (LRMs) tend to explain their reasoning, offering a promising avenue to bridge this gap. LRMs fine-tuned for chess are demonstrating consistent advancements, achieving higher Elo ratings and managing complete game sequences more effectively \cite{zhang2025complete,wang2025explore}. Optimised for step-by-step planning, LRMs possess the distinct advantage of generating explanations that resonate more intuitively with human cognitive processes \cite{sui2025fidelis}. However, the fidelity and reliability of the data generated by these models remain a challenge; LRMs are not infallible and often fail to produce faithful thinking traces, particularly under challenging conditions \cite{creswell2022faithful}. Furthermore, recent benchmarks highlight the limitations of LLMs in strategic play compared to human experts \cite{wang2025explore,zhang2024human}.

\begin{figure*}[t]
\centering
\includegraphics[width=0.90\linewidth]{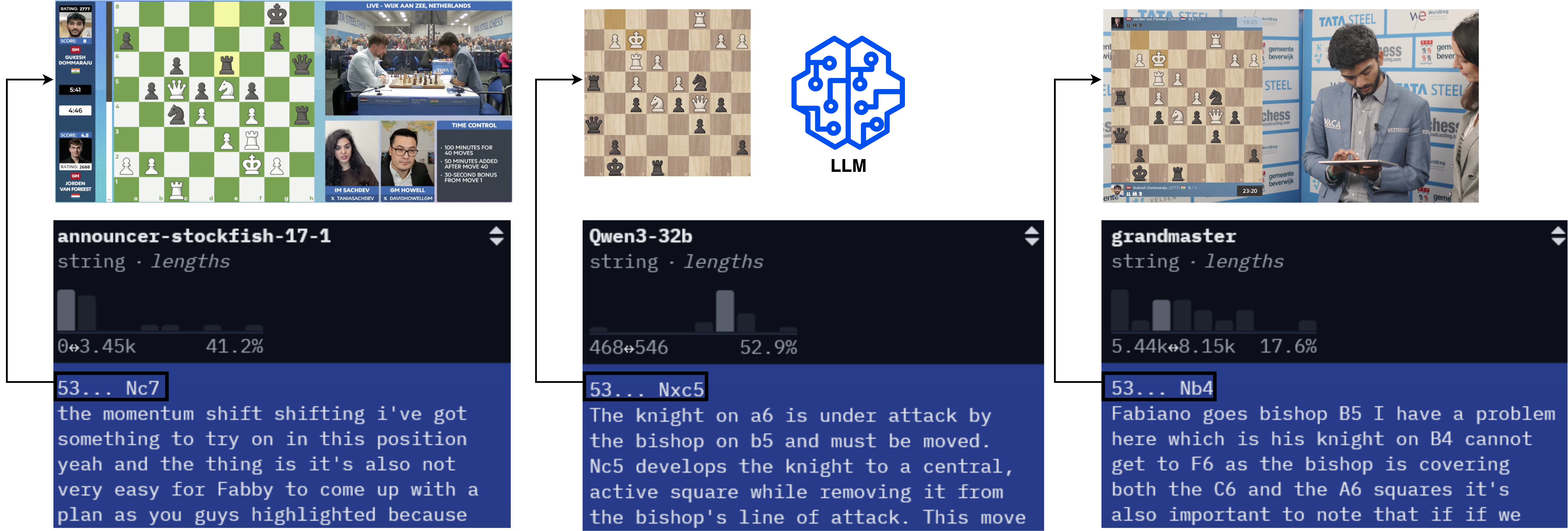}
\caption{The same chess position produces three different recommendations and explanations. From left to right: an engine-assisted commentator recommends 53\ldots Nc7, Qwen3-32B recommends 53\ldots Nxc5, and the grandmaster recommends 53\ldots Nb4. This disagreement motivates our study of alignment across engine, LLM, and human reasoning.}
\label{fig:figure1}
\end{figure*}

Chess has strict outcomes, and semantic divergences in rationale present a direct challenge for AI decision-making as a proxy for studying the safety and secure deployment of reasoning agents in higher-stakes environments. When agents with fundamentally different reasoning architectures generate semantically divergent data, how can we characterise and query this information to ensure predictable, human-aligned decision-making? We investigate this human-AI alignment through a \emph{three-body alignment} in chess. As illustrated in Figure~\ref{fig:figure1}, the reasoning behind the human experts (Grandmasters), optimisation engines (e.g., NNUEs), and generative Large Language Models (LLMs) frequently differs. They produce heterogeneous rationales ranging from intuitive moves to calculations or plausible narratives that resist conventional integration techniques and complicate the retrieval of interpretable insights.

\begin{figure*}[t]
\centering
\includegraphics[width=\textwidth]{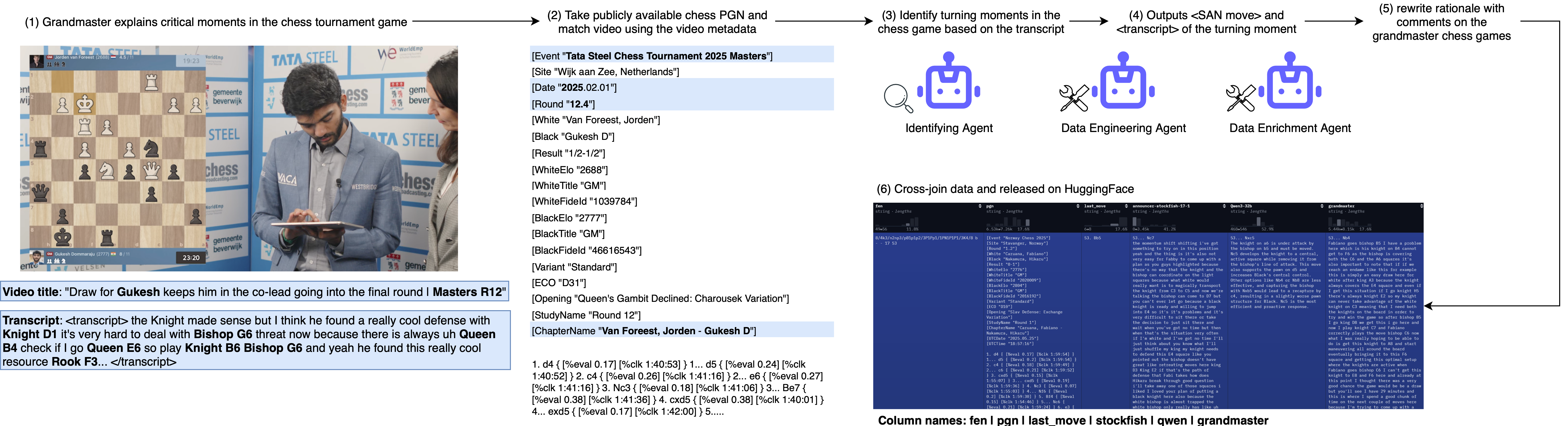}
\caption{Data resynthesis pipeline. Raw tournament video and transcripts are converted into game metadata by the identifying agent, normalised into board states and SAN moves by the data engineering agent, and augmented with engine, LLM, and grandmaster rationales by the data enrichment agent.}
\label{fig:figure2}
\end{figure*}

To address this gap and evaluate these divergent reasoning models, we introduce an agentic data resynthesis pipeline, as illustrated in Figure~\ref{fig:figure2}. This pipeline employs AI agents to convert unstructured video inputs into structured datasets. Components augmented by these agents are indicated in blue, showing the step-by-step shift from raw video data to semantically queryable formats that are amenable to subsequent alignment and security evaluations.

\paragraph{Problem Statement} We pose the following research questions (RQs) to probe the observed gap in understanding top move recommendations and agent reasoning:
\textbf{RQ1:} How can we effectively characterise and quantify the semantic divergence between fundamentally different reasoning architectures (human experts, engines, and LLMs) when evaluating strategic decisions?
\textbf{RQ2:} Can retrieval and reranking mechanisms bridge these semantic gaps to integrate divergent agent rationales into a cohesive, human-aligned framework without sacrificing performance?
\textbf{RQ3:} How can we effectively assess semantic alignment when existing benchmarks are mainly evaluated based on the presence of tokens, while standard dense retrieval techniques struggle to manage the heterogeneity of these different conceptual models?

\paragraph{Contributions} Our contributions address these RQs directly:
\textbf{C1:} A method to construct a multisource rationale database using an agentic data engineering pipeline to transform unstructured expert commentary into structured data.
\textbf{C2:} An empirical analysis of the semantic embedding space, confirming significant heterogeneity among the three reasoning sources.
\textbf{C3:} An experiment demonstrating that reranking mechanisms can improve human alignment with minimal sacrifice to performance.
\textbf{C4:} A new enriched chess narrative benchmark for evaluating text rationale similarity and exposing the limits of dense retrieval.
\textbf{C5:} The open-source release of the \texttt{trichess} dataset to support the development of secure, human-aligned intelligent systems.

\section{Related Work}

The landscape of AI research in chess spans from engines relying on exhaustive evaluation to large language models (LLMs) that integrate gameplay with strategic explanations \cite{feng2023chessgpt}. Within this research area, alignment frameworks such as MAIA-2 train on human games to emulate diverse skill levels and achieve human-like play \cite{tang2024maia,mcilroy2020learning,mcilroy2020aligning}. Amortised planning optimises strategic efficiency by pre-computing policies, blending neural networks with Monte Carlo Tree Search. While highly effective for move accuracy, these architectures rarely prioritise reasoning alignment, which limits their capacity to replicate grandmaster thought processes \cite{paes2024selective}. Recent investigations demonstrate evidence of learnt look-ahead mechanisms in chess neural networks \cite{jenner2024evidence}, and the game continues to serve as a testbed for evaluating language model state tracking \cite{toshniwal2022chess}, emergent world models \cite{karvonen2024emergent}, and the oracle approach to AI safety \cite{miller2020chess}. Studies also measure progress in dictionary learning for language model interpretability using board games \cite{karvonen2024measuring}.

Early efforts to augment engines primarily focused on natural language interfaces \cite{feng2023chessgpt}. ChessLLM, for example, achieves competitive Elo ratings through fine-tuning on complete game datasets \cite{zhang2025complete}. Despite these advances, current models overwhelmingly prioritise move correctness over coherent reasoning paths, frequently failing to capture expert thought processes \cite{zaidi2024predicting}. Our methodology bridges these domains, shifting focus toward Large Reasoning Models (LRMs) that internalise expert chess lines. As models operate increasingly as autonomous agents, ensuring their reasoning aligns with safe and predictable parameters is an urgent priority. Literature highlights the necessity of disentangling safety constraints during policy optimisation to maintain robust guardrails \cite{chua2025guardrail}. Advancements in traceable reasoning pathways, as well as long-horizon agent reasoning \cite{ye2026memweaver}, and multimodal agent recommender systems \cite{huang2025towards}, reinforce that aligning coherent reasoning pathways is just as important as winning the game through next-step prediction.

At the architectural level, retrieval mechanisms and few-shot prompting synergise well with established fine-tuning techniques. Supervised fine-tuning with Portable Game Notation (PGN) data embeds move legality and tactical patterns \cite{zhang2025complete}, while reinforcement learning from human feedback (RLHF) can steer models toward preferred strategies \cite{ouyang2022training}. The scale and quality of training data remain determinative, with contemporary models leveraging millions of human games to capture strategic diversity \cite{tang2024maia}. Efficiency optimisations like LoRA reduce the computational overhead of these processes \cite{hu2022lora}. Yet, because most approaches remain fixated on move prediction, they neglect the alignment of reasoning processes with grandmaster thought lines, yielding inconsistencies in complex tactical positions \cite{zhang2025k}. To isolate and address this misalignment, our work deliberately employs retrieval-based alignment methods. Related studies investigate goals as reward seeking models \cite{davidson2025goals} and show that aversion to external feedback stabilises agent alignment \cite{garcia2024aversion}. Underlying these systems, foundational architectures continue to evolve, integrating innovations like rotary position embeddings to better manage complex sequences \cite{su2024roformer}.

\section{Dataset}

\subsection{Nature of the Dataset}

Our dataset features chess positions from professional tournaments (e.g., Tata Steel Chess, Norway Chess) where grandmasters' games are available on Lichess. Each position is linked to a SAN move suggestion and a natural language rationale from varied sources: engines like Stockfish 17.1 (NNUE-based, explained by IM commentators), LLMs like Qwen3-32b (rationale-tagged), and grandmasters (post-game analysis). Examples show move variations (Nc7 from the NNUE engine, Nxc5 from the LLM, Nb4 from the grandmaster) for an endgame featuring a black knight on a6 attacked by a white bishop on b5, plus metadata like advantage evaluations (e.g., $+0.3$ for white) and next move probabilities (e.g., $41.2\%$). This shared-position structure makes disagreement observable at both the decision and explanation levels, rather than confounding rationale differences with different board states.

\subsection{Constructing the Data}

The pipeline transforms unstructured tournament media into replayable, position-level records through three specialised agents. Rewriting is used to normalise transcript language, while chess-aware validation preserves the connection between each rationale, its SAN move, and the corresponding board state.

\paragraph{Identifying Agent}

This agent functions as a tool-calling LLM. It processes raw video inputs (e.g., tournament clips with chessboards, player information, and transcripts) and extracts structured game metadata, including event, round, players, result, opening, and chapter identifiers. Code-based tools serialise these fields into JSON-like records, enabling game segments to be matched against PGN sources and audited before downstream enrichment.

\textit{Token Chunking.} Regarding token chunking, to address the challenge of videos containing multiple chess games, the agent employs a specialised tool that segments transcripts into smaller context windows. Using key phrases (e.g., player names, move references, or timestamps) and metadata cues to isolate and filter relevant game segments, the tool ensures only applicable portions are extracted for targeted analysis.

\paragraph{Data Engineering Agent}

This agent structures the metadata into chronological move sequences and board states, stored in tabular form with PGN, FEN, and auxiliary fields. It reconstructs positions move by move and associates transcript spans with the relevant board state. A dictionary-based mapper converts descriptive transcript phrases (e.g., ``Bishop B6'') into Standard Algebraic Notation (e.g., Bb6), enabling validation and replay with chess tooling. Illegal or ambiguous conversions can therefore be detected before they enter the rationale database.

\paragraph{Data Enrichment Agent}

This final agent integrates Stockfish evaluations, commentator explanations, grandmaster analysis, and LLM-generated rationales. It enriches each board state with source-specific text, evaluations, move context, confidence information, and candidate variations, using full PGN sequences to preserve the strategic narrative around each decision. In contrast to isolated next-move labels, the resulting records retain manoeuvres, threats, and extended lines surrounding the recommendation. This distinction is central to our use of a \emph{rationale}: it represents a source's strategic account of a decision, not merely a verbal rendering of an engine score.

\section{Methodology}

\begin{figure*}[t]
\centering
\includegraphics[width=0.78\textwidth]{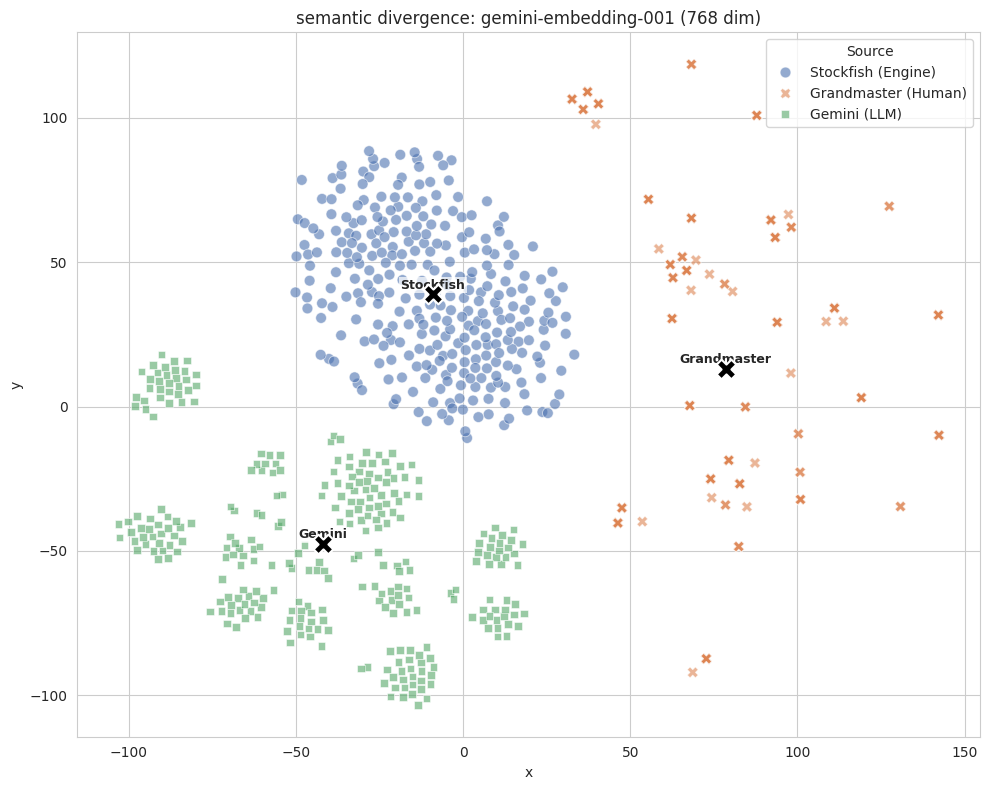}
\caption{Two-dimensional t-SNE projection of rationale embeddings from \texttt{gemini-embedding-001}. Engine-assisted commentary (blue circles), grandmaster commentary (orange crosses), and Gemini rationales (green squares) occupy visibly different regions; black crosses mark source centroids. The projection is descriptive and does not preserve all distances from the original 768-dimensional space.}
\label{fig:figure-t-SNE}
\end{figure*}

\subsection{Embedding Space Analysis via t-SNE}

To analyse the conceptual separation among the three reasoning modalities: expert intuition (grandmasters), computational precision (chess engines), and generative reasoning (LLMs), we project their natural language rationales into a shared semantic space. We utilise a dataset of chess move rationales sourced from our own constructed data from the Hugging Face repository \texttt{jaymarichua/trichess} (train split), containing columns for commentator reasoning (mapped to the chess engine, e.g., Stockfish), grandmaster reasoning (human expert), and this time select Gemini base reasoning (LLM). After loading the dataset, we filter for non-null entries in this dataset.

Each rationale is encoded using the \texttt{gemini-embedding-001} model via the Google Generative AI API, configured for clustering tasks with an output dimensionality of 768 to optimise for semantic representation in lower dimensions. Embeddings are generated in batches of 20, with text cleaned by replacing newlines to prevent parsing errors, and a 0.5-second delay between batches to adhere to rate limits. In case of API failures, fallback zero vectors are used. The resulting embeddings are normalised by dividing each vector by its L2 norm (handling zero-norm cases to avoid division errors), ensuring comparisons focus on directional similarity rather than magnitude.

The normalised embeddings from all sources are concatenated into a single matrix. We then apply t-SNE for dimensionality reduction to two components, using the following parameters: a perplexity of 30, a random state of 42, PCA initialization, and automatic learning rate adjustment. This projects the high-dimensional embeddings (768 dimensions) into a 2D space for visualisation, facilitating the inspection of cluster formations.

\paragraph{Semantic Divergence in Reasoning Embeddings}
The visualisation maps the distinct latent spaces occupied by Stockfish engine evaluations (blue), Grandmaster human commentary (orange), and Gemini LLM generated reasoning (green). This projection highlights the separation among the three reasoning paradigms. Stockfish outputs form a dense cluster, reflecting the high consistency and low semantic entropy of algorithmic evaluations. Conversely, Grandmaster analysis exhibits significant spatial dispersion, indicative of the varied, narrative, and highly contextual nature of human explanation. Notably, the Gemini LLM does not function as an intermediary between these human and engine baselines; instead, it isolates itself in a distinct region of the embedding space. The model's outputs organise into several tight clusters, suggesting it operates within unique modes of synthetic reasoning that are semantically distinguishable from both biological intuition and pure calculation.

We create a scatterplot with points coloured by source: Stockfish-assisted commentary in blue, grandmaster commentary in orange, and Gemini reasoning in green. To provide visual reference points, we compute the mean 2D coordinate for each source and mark it with a large black cross. Rigorous quantification of semantic drift is performed later using cosine similarity in the original high-dimensional space; the t-SNE plot in Figure~\ref{fig:figure-t-SNE} is used only for exploratory visualisation.

The results reveal distinct clustering by source, with grandmaster rationales often grouping around strategic concepts (e.g., ``initiative,'' ``weakness,'' and ``prophylaxis''), engine rationales emphasising tactical precision, and LLM rationales showing varied generative patterns. This separation supports the need for alignment mechanisms that can mediate between distinct cognitive and linguistic paradigms.

\subsection{One-Shot Source Classification}

For each test rationale, we provide in-context examples per class from one source (e.g., a grandmaster annotation) and prompt the model to classify the target rationale into one of the three distinct source categories. Uniquely, we enrich the prompt with the contextual game state to test whether the model utilises actual board dynamics rather than just surface-level linguistic cues to classify the reasoning source. In providing this game-aware setting, we were able to take this further and prompt the model to return the top-\emph{k} diagnostic terms used in its decision (e.g., "outpost," "calculation depth," "simplification").

Aggregating responses over 500 randomly sampled entries, we observe that accuracy improves from 63\% (blind) to 72\% (game-aware), significantly above chance (33\%). Notably, in the game-aware setting, the model increasingly relies on strategic coherence (e.g., "punishes f5," "central breakthrough") rather than spurious surface-level cues.

\begin{figure*}[t]
    \centering

    \begin{subfigure}{\linewidth}
        \centering
        \includegraphics[trim=0 0 575bp 0,clip,width=0.664\linewidth]{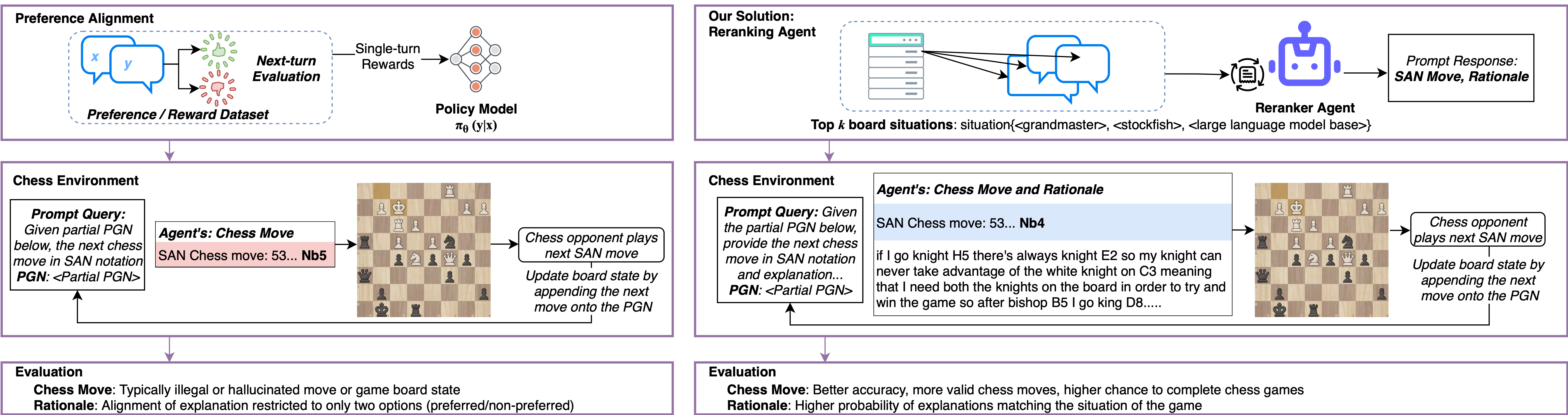}
        \caption{Conventional preference optimisation: pairwise rewards train a policy that directly outputs the next move.}
        \label{fig:workflow_preference}
    \end{subfigure}

    \vspace{2ex}

    \begin{subfigure}{\linewidth}
        \centering
        \includegraphics[trim=434bp 0 0 0,clip,width=0.88\linewidth]{figure3-x.png}
        \caption{Proposed RR-RAG workflow: position-specific evidence is retrieved and reranked before the model generates a SAN move and rationale.}
        \label{fig:workflow_rrrag}
    \end{subfigure}

    \caption{Conceptual comparison of two chess-agent workflows. The panels are cropped from the original side-by-side diagram and stacked vertically so that the annotations and chessboards remain legible. Panel~(a) provides architectural context rather than an evaluated RLHF baseline; our experiments compare panel~(b) with zero-shot and standard RAG baselines.}
    \label{fig:example3}
\end{figure*}

\subsection{Rationale Reranking for Retrieval Augmented Generation, RR-RAG}

The classification results demonstrate that language models can distinguish between reasoning architectures, particularly when grounded in the actual game state. Building directly on this finding, we hypothesise that explicitly curating these sources and filtering them by board dynamics can compel a generative model to adopt a specific, human-aligned reasoning style. Thus, we implement retrieval-based reranking designed to generate instructive moves. Given a chess position (FEN), the system proceeds in two distinct stages.

In the first stage, dense retrieval, we retrieve k=10 candidate move-rationale pairs from our annotated knowledge base using dense semantic search. Embeddings are computed via the standard sentence transformer, and retrieval is performed using the FAISS vector database \cite{douze2025faiss}. Candidates are drawn from all three sources: grandmaster annotations, expert commentaries, and the LRM's explanations.


In the second stage, we apply a heuristic contextual filtering mechanism. Recognising that relying solely on text-to-text semantic search is insufficient for chess strategy, we compute the Inner Product (cosine similarity) between the text embeddings of the current and historical FEN strings using FAISS. We use this strictly as a string-matching proxy to narrow down structurally similar candidate boards. We acknowledge a fundamental limitation here: standard dense language models lack native spatial awareness of FEN geometry, meaning minor character changes can drastically alter the objective board evaluation despite high embedding similarity. Future work should replace text embeddings with character-level string distance metrics (e.g., Levenshtein distance) or piece-overlap evaluations to better capture spatial board geometry. Nonetheless, our baseline isolates the top three candidate positions to retrieve their associated Grandmaster rationales.

\section{Experiments}\label{sec:experiments}

Since solving our research problem lies in resolving the semantic divergence among heterogeneous reasoning sources, our experiment is set up to test whether a retrieval-augmented reranking framework can successfully align a reasoning model, i.e., \emph{Gemini 3 Flash (gemini-3-flash-preview) and Gemini 3.1 Pro (gemini-3.1-pro-preview)}, with grandmaster intuition and investigate any trade-offs from this forced alignment.

\subsection{Experimental Setup}

We set up a reliable baseline of expert data such that our evaluation uses a dataset of 760 FEN-rationale pairs extracted from professional tournament games. These natural language rationales are embedded using the \texttt{gemini-embedding-001} model and indexed via FAISS-GPU to enable efficient semantic retrieval. Testing the model on tournament game positions dictated the construction of our test set, which comprises 200 distinct positions generated by parsing and truncating PGN sequences at critical junctures. These truncations typically occur using the identifying agent when there is a relatively high change in evaluation scores of the total move sequence or immediately following annotated blunder positions. Subjecting these positions to minor perturbations, such as targeted piece swaps inspired by described inaccuracies (e.g., modifying a board state to reflect the blunder 53...Bg6?), forces the model to reason through entirely new board states.

Strategic reasoning for each test position is generated under three distinct experimental conditions. The zero-shot baseline relies entirely on the language model's internal weights and remains devoid of external retrieved rationales. The Retrieval-Augmented Generation (RAG) condition introduces a standard few-shot approach that augments the prompt with the top-3 unrefined candidates selected from $k=10$ retrieved items. Our proposed Rationale-Reranked RAG (RR-RAG) expands this retrieval pool to $k=20$ and applies the aforementioned FEN-based inner product filtering mechanism to isolate and rank the final top-3 most contextually relevant candidates. We also assess the computational trade-offs of model scale, which involves comparing two variants for the final generation step: the efficient \texttt{gemini-3-flash} and the more reasoning-intensive \texttt{gemini-3-pro}.

\subsection{Evaluation Metrics}

Our design for performance measurement addresses both strategic style alignment and tactical correctness. To quantify style alignment, we establish a position-specific Grandmaster baseline. Style alignment is then measured strictly as the cosine similarity between the embedding of the language model's generated rationale and the specific grandmaster rationale for that exact board position, rather than a generic global centroid.

For tactical proficiency, objective move quality is evaluated by calculating the centipawn loss (regret) between the engine's optimal move and the generated move using Stockfish 17.1. To standardise these evaluations, we apply an exponential decay function to the loss ($e^{-0.005 \times \text{loss}}$). This normalises the tactical score on a scale from 0.0 (blunder or hallucinated move) to 1.0 (engine-optimal), allowing for stable comparisons across varying degrees of move inaccuracy.

\section{Results}

\begin{figure*}[t]
    \centering
    \begin{minipage}[t]{0.49\textwidth}
        \centering
        \includegraphics[width=\linewidth]{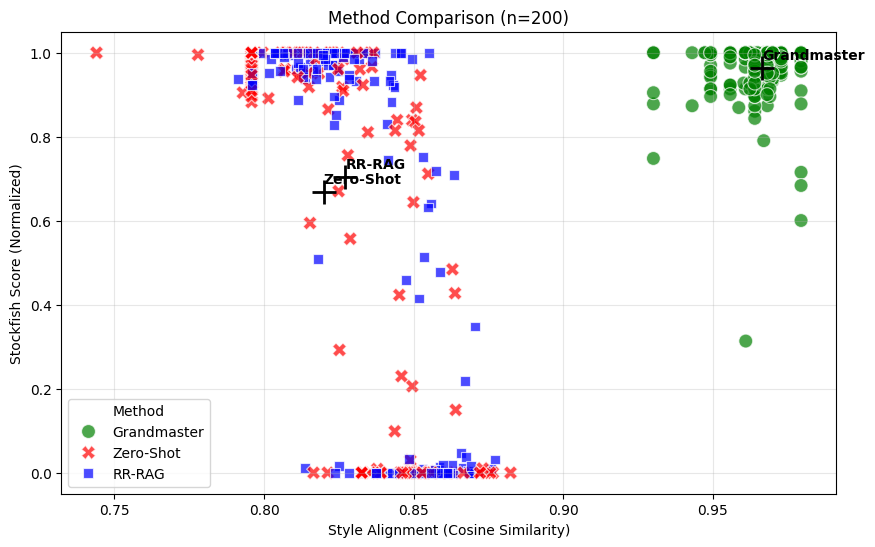}
        \small (a) Tactical quality versus human alignment
    \end{minipage}
    \hfill
    \begin{minipage}[t]{0.49\textwidth}
        \centering
        \includegraphics[width=\linewidth]{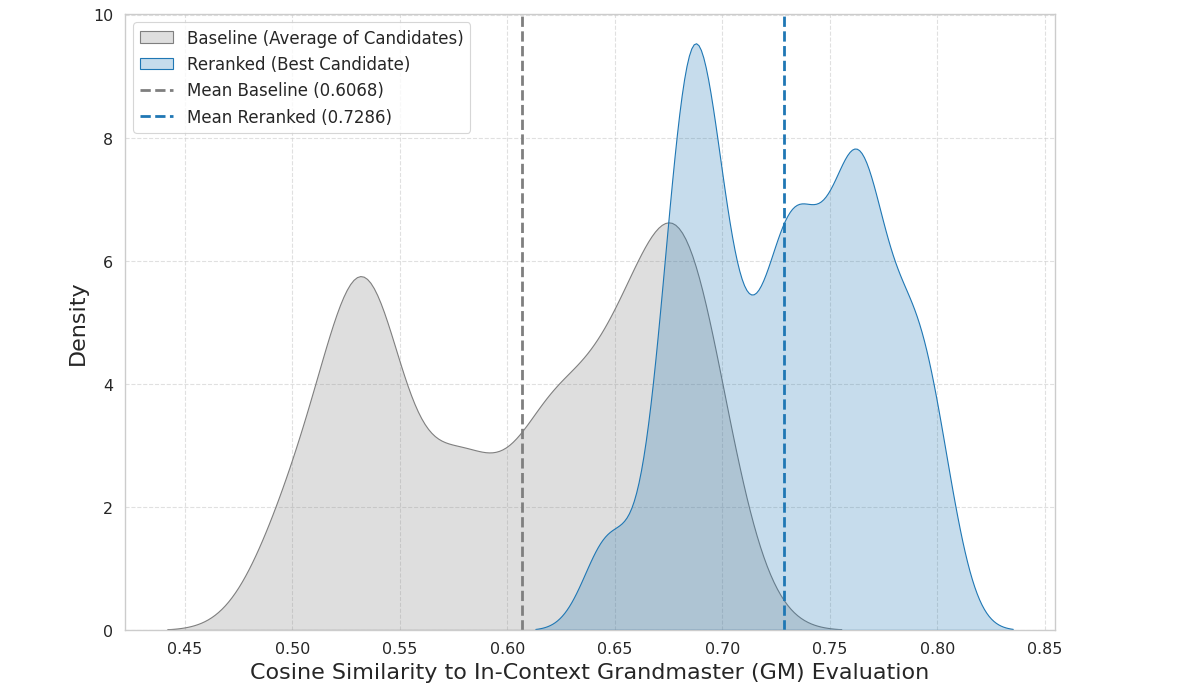}
        \small (b) Distribution of rationale alignment scores
    \end{minipage}
    \caption{Alignment results on 200 test positions. (a) Generated moves are compared by normalised Stockfish score and cosine similarity to the position-specific grandmaster rationale. (b) Reranking shifts the mean rationale similarity from 0.6068 for standard RAG to 0.7286 for RR-RAG.}
    \label{fig:alignment_results}
\end{figure*}

The inherent tension between computational optimality and human-relatable strategy becomes immediately visible when mapping the generated rationales. Plotting style alignment against normalised Stockfish scores (Figure~\ref{fig:alignment_results}a) reveals a stark distributional shift. The Grandmaster baseline naturally clusters in the upper-right quadrant, occupying the ideal intersection of high tactical accuracy and deep stylistic alignment. In contrast, both the Zero-Shot and RR-RAG methods achieve high tactical scores but remain noticeably separated from the human baseline in style alignment.

Applying the FEN-based contextual filtering mechanism in RR-RAG actively bridges this gap. The standard RAG baseline, which measures the semantic alignment of the model's generated rationale when conditioned on unrefined retrieved candidates, yields a moderate mean alignment of 0.6068. In contrast, the RR-RAG setup actively conditions the model on curated grandmaster reasoning, inducing a definitive rightward shift in the distribution over standard RAG (Figure~\ref{fig:alignment_results}b). The reranked outputs concentrate significantly higher and achieve a mean alignment of 0.7286. This absolute improvement of +0.1218 translates to a 20.1\% relative gain in alignment with grandmaster intuition. The effect is particularly pronounced in perturbed test positions, where evaluating lines such as ['Qd6'] or navigating blunders like 53...Bg6? requires the model to prioritize strategic depth and long-term initiative over rigid computational precision.

\begin{figure*}[t]
    \centering
    \begin{minipage}[b]{0.48\textwidth}
        \centering
        \includegraphics[width=\linewidth]{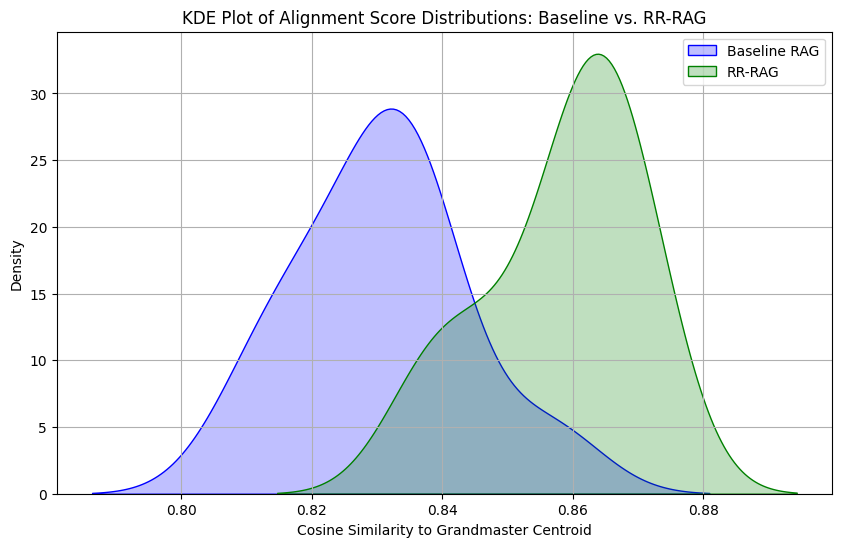}
        \vspace{0.1cm} 
        \centerline{(a) RR-RAG w/ Gemini Flash}
    \end{minipage}
    \hfill
    \begin{minipage}[b]{0.48\textwidth}
        \centering
        \includegraphics[width=\linewidth]{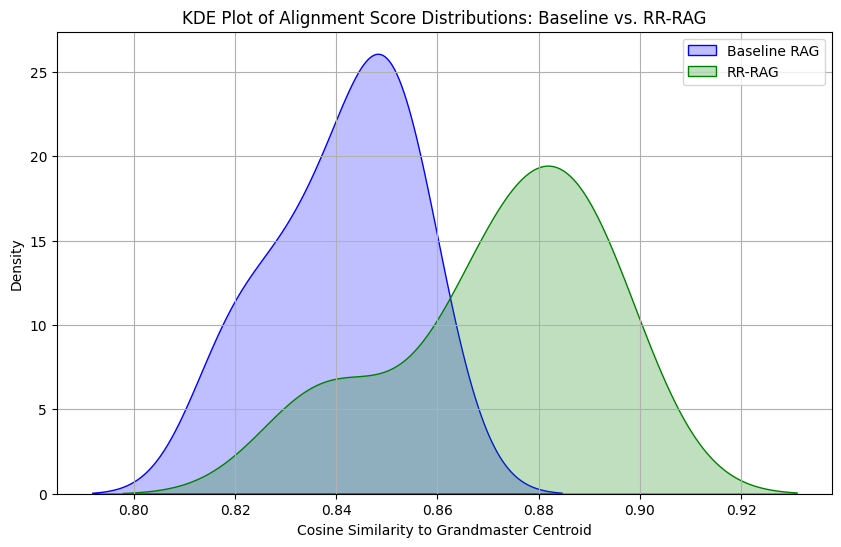}
        \vspace{0.1cm}
        \centerline{(b) RR-RAG w/ Gemini Pro}
    \end{minipage}

    \caption{RR-RAG response distributions for (a) Gemini Flash and (b) Gemini Pro. The Pro distribution is narrower than the Flash distribution, indicating lower observed variation across generations in this experiment.}
    \label{fig:gemini_comparison}
\end{figure*}

We also find that the scale of the underlying reasoning model affects generation stability. The \texttt{gemini-3-pro} model exhibits a markedly narrower distribution than \texttt{gemini-3-flash} (Figure~\ref{fig:gemini_comparison}). This reduction in variance suggests that the Pro model more reliably converges on consistent strategic outputs, suppressing some of the localised volatility observed in the lighter model. Because the comparison is observational, we interpret the narrower band as evidence of greater consistency in this experiment rather than attributing it solely to model size.

\section{Discussion}

Resolving semantic divergence by prioritising grandmaster-style outputs carries a cost. RR-RAG raises mean alignment to 0.73 from a 0.50 zero-shot baseline and lifts normalised tactical quality from 0.40 to 0.75, but the boundary conditions expose the deeper complexity of the three-body problem.

Alignment introduces an explicit performance tax: aggressively conditioning the model on narrative sources forces a trade-off in which maximising human-style reasoning can reduce engine-evaluated move quality. Peak normalised tactical scores decline from 0.8 to 0.6 in positions that favour human-like strategic foresight over algorithmic calculation. This tax also appears within the retrieval architecture: RR-RAG reaches alignment scores of 0.75, but sacrifices up to 0.3 on the normalised Stockfish scale relative to unconstrained few-shot RAG. Thus, improved semantic agreement should not be interpreted as evidence that the selected move is objectively stronger.

Strict retrieval-augmented alignment can therefore amplify the misalignment it intends to solve. Compelling a system to adopt a single reasoning geometry on whether grandmaster narrative or engine calculation reduces the flexibility required for non-standard but optimal lines and may entrench strategic blind spots. Reconciling multisource knowledge requires more than richer human data: it requires intermediate mechanisms that mediate between human-centred explanations and the objective accuracy of NNUE engines rather than optimising either source in isolation.

\section{Future Work}

The limitations encountered in evaluation highlight the need to move beyond simply mimicking grandmaster narratives and towards alignment with absolute, ground-truth solutions. To support this next phase, we developed a more demanding benchmark based on chess puzzles. Standard puzzle datasets are abundant, but generally lack the detailed tactical explanations required to train and evaluate complex reasoning systems. We therefore extend \textit{trichess} with an enriched solution structure that maps key moves and tactical motifs to step-by-step reasoning (Figure~\ref{fig:figure4}). Although comprehensive model evaluation on this benchmark is reserved for future work, these narrative puzzles provide concrete training examples for generating interpretable, competitive chess strategies and for testing whether semantic similarity corresponds to tactically faithful reasoning.

\begin{center}
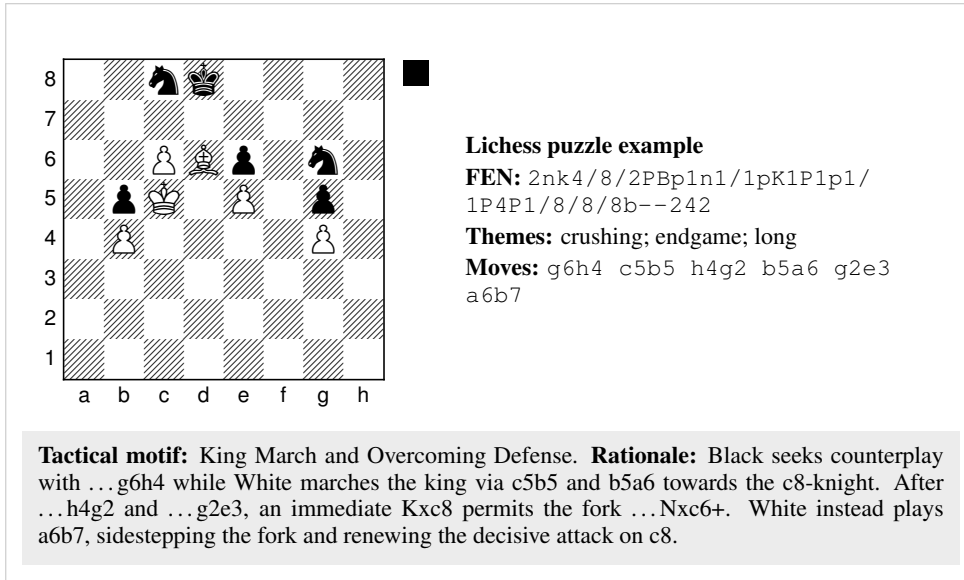

\begingroup
\setlength{\fboxsep}{6pt}
\fcolorbox{gray!45}{white}{%
\begin{minipage}{\dimexpr0.92\linewidth-2\fboxsep-2\fboxrule\relax}
\setlength{\parindent}{0pt}
\begin{minipage}[c]{0.43\linewidth}
\centering
\chessboard[smallboard, setfen=2nk4/8/2PBp1n1/1pK1P1p1/1P4P1/8/8/8 b - - 2 42]
\end{minipage}%
\hfill
\begin{minipage}[c]{0.52\linewidth}
\raggedright\small
\textbf{Lichess puzzle example} \\[0.5ex]
\textbf{FEN:} \nolinkurl{2nk4/8/2PBp1n1/1pK1P1p1/1P4P1/8/8/8 b - - 2 42} \\[0.5ex]
\textbf{Themes:} crushing; endgame; long \\[0.5ex]
\textbf{Moves:} \texttt{g6h4 c5b5 h4g2 b5a6 g2e3 a6b7}
\end{minipage}
\vspace{0.8ex}

\colorbox{gray!15}{%
\begin{minipage}{\dimexpr\linewidth-2\fboxsep\relax}
\footnotesize
\textbf{Tactical motif:} King March and Overcoming Defense. 
\textbf{Rationale:} Black seeks counterplay with \ldots g6h4 while White marches the king via c5b5 and b5a6 towards the c8-knight. After \ldots h4g2 and \ldots g2e3, an immediate Kxc8 permits the fork \ldots Nxc6+. White instead plays a6b7, sidestepping the fork and renewing the decisive attack on c8.
\end{minipage}}
\end{minipage}}
\endgroup

\captionof{figure}{An enriched puzzle record combining the board, machine-readable solution, tactical motif, and grounded narrative.}
\label{fig:figure4}
\end{center}

\section{Conclusion}

We frame the semantic divergence among human experts, optimisation engines, and language models as the three-body alignment problem. Our experiments show that RR-RAG can shift generated rationales towards grandmaster explanations and improve mean semantic alignment. However, this improvement incurs a measurable loss in engine-evaluated move quality, confirming that current retrieval mechanisms cannot easily maximise human interpretability and tactical precision simultaneously. Rather than treating this trade-off solely as a system failure, we establish the alignment sacrifice as a concrete evaluation target for future models. The released \textit{trichess} resources and enriched puzzle benchmark provide a novel empirical technique for developing semantic architectures that reconcile conflicting knowledge sources, ground explanations in verifiable tactical lines, and preserve correct, human-centred decisions.

\bibliographystyle{iclr2026_conference}
\bibliography{ss}

@article{tang2024maia,
  title={Maia-2: A unified model for human-ai alignment in chess},
  author={Tang, Zhenwei and Jiao, Difan and McIlroy-Young, Reid and Kleinberg, Jon and Sen, Siddhartha and Anderson, Ashton},
  journal={Advances in Neural Information Processing Systems},
  volume={37},
  pages={20919--20944},
  year={2024}
}

@article{ruoss2024amortized,
  title={Amortized planning with large-scale transformers: A case study on chess},
  author={Ruoss, Anian and Del{\'e}tang, Gr{\'e}goire and Medapati, Sourabh and Grau-Moya, Jordi and Wenliang, Li K and Catt, Elliot and Reid, John and Lewis, Cannada A and Veness, Joel and Genewein, Tim},
  journal={Advances in Neural Information Processing Systems},
  volume={37},
  pages={65765--65790},
  year={2024}
}

@inproceedings{zhang2025complete,
  title={Complete chess games enable llm become a chess master},
  author={Zhang, Yinqi and Han, Xintian and Li, Haolong and Chen, Kedi and Lin, Shaohui},
  booktitle={Proceedings of the 2025 Conference of the Nations of the Americas Chapter of the Association for Computational Linguistics: Human Language Technologies (Volume 2: Short Papers)},
  pages={1--7},
  year={2025}
}

@article{zhang2024human,
  title={Human-aligned chess with a bit of search},
  author={Zhang, Yiming and Jacob, Athul Paul and Lai, Vivian and Fried, Daniel and Ippolito, Daphne},
  journal={arXiv preprint arXiv:2410.03893},
  year={2024}
}

@inproceedings{wang2025explore,
  title={Explore the reasoning capability of llms in the chess testbed},
  author={Wang, Shu and Ji, Lei and Wang, Renxi and Zhao, Wenxiao and Liu, Haokun and Hou, Yifan and Wu, Ying Nian},
  booktitle={Proceedings of the 2025 Conference of the Nations of the Americas Chapter of the Association for Computational Linguistics: Human Language Technologies (Volume 2: Short Papers)},
  pages={611--622},
  year={2025}
}

@inproceedings{sui2025fidelis,
  title={Fidelis: Faithful reasoning in large language models for knowledge graph question answering},
  author={Sui, Yuan and He, Yufei and Liu, Nian and He, Xiaoxin and Wang, Kun and Hooi, Bryan},
  booktitle={Findings of the Association for Computational Linguistics: ACL 2025},
  pages={8315--8330},
  year={2025}
}

@misc{creswell2022faithful,
  title={Faithful Reasoning Using Large Language Models},
  author={Antonia Creswell},
  year={2022},
  eprint={2208.14271},
  archivePrefix={arXiv},
  primaryClass={cs.AI}
}

@article{feng2023chessgpt,
  title={Chessgpt: Bridging policy learning and language modeling},
  author={Feng, Xidong and Luo, Yicheng and Wang, Ziyan and Tang, Hongrui and Yang, Mengyue and Shao, Kun and Mguni, David and Du, Yali and Wang, Jun},
  journal={Advances in Neural Information Processing Systems},
  volume={36},
  pages={7216--7262},
  year={2023}
}

@article{mcilroy2020learning,
  title={Learning personalized models of human behavior in chess},
  author={McIlroy-Young, Reid and Wang, Russell and Sen, Siddhartha and Kleinberg, Jon and Anderson, Ashton},
  journal={arXiv preprint arXiv:2008.10086},
  year={2020}
}

@inproceedings{mcilroy2020aligning,
  title={Aligning superhuman ai with human behavior: Chess as a model system},
  author={McIlroy-Young, Reid and Sen, Siddhartha and Kleinberg, Jon and Anderson, Ashton},
  booktitle={Proceedings of the 26th ACM SIGKDD international conference on knowledge discovery \& data mining},
  pages={1677--1687},
  year={2020}
}

@article{paes2024selective,
  title={Selective explanations},
  author={Paes, Lucas M and Wei, Dennis and Calmon, Flavio P},
  journal={Advances in Neural Information Processing Systems},
  volume={37},
  pages={55562--55590},
  year={2024}
}

@article{jenner2024evidence,
  title={Evidence of learned look-ahead in a chess-playing neural network},
  author={Jenner, Erik and Kapur, Shreyas and Georgiev, Vasil and Allen, Cameron and Emmons, Scott and Russell, Stuart},
  journal={Advances in Neural Information Processing Systems},
  volume={37},
  pages={31410--31437},
  year={2024}
}

@inproceedings{toshniwal2022chess,
  title={Chess as a testbed for language model state tracking},
  author={Toshniwal, Shubham and Wiseman, Sam and Livescu, Karen and Gimpel, Kevin},
  booktitle={Proceedings of the AAAI Conference on Artificial Intelligence},
  volume={36},
  number={10},
  pages={11385--11393},
  year={2022}
}

@article{karvonen2024emergent,
  title={Emergent world models and latent variable estimation in chess-playing language models},
  author={Karvonen, Adam},
  journal={arXiv preprint arXiv:2403.15498},
  year={2024}
}

@article{miller2020chess,
  title={Chess as a testing grounds for the oracle approach to AI safety},
  author={Miller, James D and Yampolskiy, Roman and Haggstrom, Olle and Armstrong, Stuart},
  journal={arXiv preprint arXiv:2010.02911},
  year={2020}
}

@article{karvonen2024measuring,
  title={Measuring progress in dictionary learning for language model interpretability with board game models},
  author={Karvonen, Adam and Wright, Benjamin and Rager, Can and Angell, Rico and Brinkmann, Jannik and Smith, Logan and Mayrink Verdun, Claudio and Bau, David and Marks, Samuel},
  journal={Advances in Neural Information Processing Systems},
  volume={37},
  pages={83091--83118},
  year={2024}
}

@article{zaidi2024predicting,
  title={Predicting user perception of move brilliance in chess},
  author={Zaidi, Kamron and Guerzhoy, Michael},
  journal={arXiv preprint arXiv:2406.11895},
  year={2024}
}

@inproceedings{chua2025guardrail,
  title={Guardrail Guided Policy Optimisation: Learning Disentangled Safety Constraints},
  author={Chua, Jaymari and Wang, Chen and Zhu, Liming and Yao, Lina},
  booktitle={Australasian Joint Conference on Artificial Intelligence},
  pages={414--425},
  year={2025},
  organization={Springer}
}

@article{ye2026memweaver,
  title={MemWeaver: Weaving Hybrid Memories for Traceable Long-Horizon Agentic Reasoning},
  author={Ye, Juexiang and Li, Xue and Yang, Xinyu and Huang, Chengkai and Nie, Lanshun and Yao, Lina and Zhan, Dechen},
  journal={arXiv preprint arXiv:2601.18204},
  year={2026}
}

@article{huang2025towards,
  title={Towards agentic recommender systems in the era of multimodal large language models},
  author={Huang, Chengkai and Wu, Junda and Xia, Yu and Yu, Zixu and Wang, Ruhan and Yu, Tong and Zhang, Ruiyi and Rossi, Ryan A and Kveton, Branislav and Zhou, Dongruo and others},
  journal={arXiv preprint arXiv:2503.16734},
  year={2025}
}

@article{ouyang2022training,
  title={Training language models to follow instructions with human feedback},
  author={Ouyang, Long and Wu, Jeffrey and Jiang, Xu and Almeida, Diogo and Wainwright, Carroll and Mishkin, Pamela and Zhang, Chong and Agarwal, Sandhini and Slama, Katarina and Ray, Alex and others},
  journal={Advances in neural information processing systems},
  volume={35},
  pages={27730--27744},
  year={2022}
}

@article{hu2022lora,
  title={Lora: Low-rank adaptation of large language models.},
  author={Hu, Edward J and Shen, Yelong and Wallis, Phillip and Allen-Zhu, Zeyuan and Li, Yuanzhi and Wang, Shean and Wang, Liang and Chen, Weizhu and others},
  journal={Iclr},
  volume={1},
  number={2},
  pages={3},
  year={2022}
}

@inproceedings{zhang2025k,
  title={K-Level Reasoning: Establishing Higher Order Beliefs in Large Language Models for Strategic Reasoning},
  author={Zhang, Yadong and Mao, Shaoguang and Ge, Tao and Wang, Xun and Xia, Yan and Lan, Man and Wei, Furu},
  booktitle={Proceedings of the 2025 Conference of the Nations of the Americas Chapter of the Association for Computational Linguistics: Human Language Technologies (Volume 1: Long Papers)},
  pages={7212--7234},
  year={2025}
}

@article{davidson2025goals,
  title={Goals as reward-producing programs},
  author={Davidson, Guy and Todd, Graham and Togelius, Julian and Gureckis, Todd M and Lake, Brenden M},
  journal={Nature Machine Intelligence},
  volume={7},
  number={2},
  pages={205--220},
  year={2025},
  publisher={Nature Publishing Group UK London}
}

@article{garcia2024aversion,
  title={Aversion to external feedback suffices to ensure agent alignment},
  author={Garcia, Paulo},
  journal={Scientific Reports},
  volume={14},
  number={1},
  pages={21147},
  year={2024},
  publisher={Nature Publishing Group UK London}
}

@article{su2024roformer,
  title={Roformer: Enhanced transformer with rotary position embedding},
  author={Su, Jianlin and Ahmed, Murtadha and Lu, Yu and Pan, Shengfeng and Bo, Wen and Liu, Yunfeng},
  journal={Neurocomputing},
  volume={568},
  pages={127063},
  year={2024},
  publisher={Elsevier}
}

@article{douze2025faiss,
  title={The faiss library},
  author={Douze, Matthijs and Guzhva, Alexandr and Deng, Chengqi and Johnson, Jeff and Szilvasy, Gergely and Mazar{\'e}, Pierre-Emmanuel and Lomeli, Maria and Hosseini, Lucas and J{\'e}gou, Herv{\'e}},
  journal={IEEE Transactions on Big Data},
  year={2025},
  publisher={IEEE}
}






\end{document}